\documentclass[aps,pra,amsmath,amssymb,twocolumn,showpacs,showkeys]{revtex4}
\usepackage{amsmath,amsthm,amssymb}
\usepackage{graphicx}
\usepackage{pstricks}

\newcommand{\ci}[1]{\cite{#1}}

\begin{document}

\title {Dirac Particles in Transparent Quantum  Graphs: \\
Tunable transport of relativistic quasiparticles in branched structures}
\author{J.R. Yusupov$^1$, K.K. Sabirov$^{2}$, Q.U. Asadov$^{2}$, M. Ehrhardt$^{3}$ and D.U. Matrasulov$^1$}
\affiliation{$^1$Turin Polytechnic University in Tashkent, 17
Niyazov Str., 100095, Tashkent, Uzbekistan\\
$^2$Tashkent University of Information Technologies, 108 Amir
Temur Str., 100200, Tashkent Uzbekistan\\
$^3$Bergische Universit\"at Wuppertal, Gau{\ss}strasse 20, D-42119
Wuppertal, Germany}

\begin{abstract}
We consider the dynamics of relativistic spin-half particles in quantum graphs
with transparent branching points. The system is modeled by combining the
quantum graph concept with the one of transparent boundary conditions applied
to the Dirac equation on metric graphs. Within such an approach, we derive
simple constraints, which turn the usual Kirchhoff-type boundary conditions at the vertex equivalent to the transparent ones. Our method is applied to quantum star graph. 
An extension to more complicated graph topologies is straightforward.
\end{abstract}


\maketitle

\section{Introduction}

Due to the recent development in condensed matter physics, modeling of
relativistic particle dynamics in low-dimensional systems attracted much
attention. It has been discovered that such materials as graphene
\ci{Neto,Kotov,BF}, carbon nanotubes (CNT) \ci{CNT}, topological insulators
\ci{TI1,TI2}, and some types of superconductors \ci{Jackiw,Alicea1}, can
provide quasiparticle excitations, which can mimic relativistic dynamics. Such
quasiparticles are described in terms of the Dirac, Weyl and Bogoliubov de
Gennes equations. Low dimensional functional materials, in which such
quasiparticles appear, can be used for engineering of different nanoscale
electronic and optoelectronic devices in emerging quantum technologies.

Effective functionalization and device optimization in such technologies
require solving the problem of tunable particle transport in low-dimensional
structures. Subsequently, this requires developing effective and realistic
models for the relativistic quasiparticle transport in low-dimensional systems.
As most of the low-dimensional systems, arising in condensed matter, exhibit a
branched structure, the above task can be reduced to the problem of
relativistic quasiparticle dynamics modeling in branched quantum systems. These
latter mentioned are usually modeled in terms of so-called quantum graphs,
which has attracted much attention during past three decades (see,
Refs.~\ci{Uzy1,Hul,Kuchment04,Uzy2,Exner15} for review recent developments in
this research field).

In this paper we consider the problem of transparent quantum graphs for
relativistic (spin-half) quasiparticles described in terms of the Dirac
equation on metric graphs. The model we propose can be directly applied to the
problem of tunable transport of relativistic quasiparticles in branched
graphene nanoribbons, carbon nanoribbon networks, topological insulator
networks and branched optical fibers, where the Dirac quasiparticles appear.
Combining quantum graph and transparent boundary conditions concepts, we derive
explicit vertex boundary conditions, providing reflectionless transport of
quasiparticles through the branching points. We reveal simple constraints for
the regime, when the motion of relativistic spin-half quasiparticle in quantum
graph becomes reflectionless, i.e., the quantum graph becomes transparent with
respect to the particle transport.

The motivation for the study of transparent quantum graphs arises from the wide
range of practical applications in optoelectronics and condensed matter
physics. Effective signal transfer in optoelectronic networks requires a
reduction to minimum losses. Spin, charge and energy transport in branched
nanomaterials used in nanoelectronic devices also requires minimization of
losses caused by backscattering of quasiparticles. In the context of
relativistic quantum dynamics with reflectionless transport, this problem will
appear in low-dimensional functional materials (e.g., graphene, CNT,
topological insulators, one-dimensional branched quasimolecules, etc).
Different quantum wire networks appearing in solid state physics are also
potential structures, where the wave transport can be tuned from the diffusive to
the ballistic regime using transparent boundary conditions.

The paper is organized as follows. In the next section we give a brief
description of the transparent boundary conditions for the Dirac equation on a real
line. Section~III presents a short introduction to the Dirac equation on quantum
graphs. Section~IV includes our study of Dirac particles in transparent quantum
graphs. A numerical justification of the obtained results is stated in
Section~V. Finally, Section~VI concludes our work with some final remarks.

\section{Transparent boundary conditions for the Dirac equation on a line}

The standard way for the description of particle and wave scattering in quantum
mechanics is the scattering matrix based approach. However, within such an
approach one does not use explicit solutions of the Schr\"odinger equation and
hence, it is less effective for the cases, when one needs to describe
the transmission of the wave through the boundary of two domains. 
Moreover, it does not provide any solution for the problem of tunable scattering and transmission
of particles for a given point/domain using the initial conditions for the
Schr\"odinger equation.

An effective approach providing solutions of the problem of reflectionless
transmission of the waves and particles through the given point/boundary in
terms of a Cauchy problem has attracted much attention during the past three
decades. Within such an approach, one can formulate boundary conditions for the
absence of backscattering at a given point or domain boundary, although the
explicit form of such boundary conditions are much more complicated than those
of Dirichlet, Neumann and Robin conditions. Such boundary conditions are called
transparent boundary conditions (TBCs).

A TBC is a boundary condition for a wave-type equation, 
which leads to a solution providing minimum (or absence) of reflections. 
In early works (see, e.g., \ci{EM1,EM2,Halpern,Sofronov,Shibata,Kuska}) 
the TBC was defined as a boundary conditions providing a minimum of backscattering in terms of the reflection coefficient. 
However, later a more strict mathematical definition of a TBC has been provided as follows \cite{Arnold1998,Ehrhardt1999,Ehrhardt2001,Ehrhardt2002,Arnold2003,Jiang2004,Antoine2008,Ehrhardt2008,Sumichrast2009,Antoine2009,Ehrhardt2010,Klein2011,Arnold2012,Feshchenko2013,Antoine2014}. 
For a given finite domain, $\Omega$, the TBCs are imposed in such a way that the solution of a PDE in $\Omega$ 
corresponds to that in the whole space, i.e., the wave/particle moving inside/outside the domain does not `see' the boundary of the domain. 
The construction of the TBC is based on the coupling of initial value boundary problems (IVBPs) in interior and exterior domains \cite{Arnold1998,Ehrhardt1999,Ehrhardt2001,Ehrhardt2002,Arnold2003,Jiang2004,Antoine2008,Ehrhardt2008,Sumichrast2009,Antoine2009,Ehrhardt2010,Klein2011,Arnold2012,Feshchenko2013,Antoine2014}.

The general procedure for constructing transparent boundary conditions on a real line here reads as follows, cf.~\cite{Antoine2008}
\begin{enumerate}\setlength{\itemsep}{0cm}
    \item Split the original PDE evolution problem into coupled equations: the interior and exterior problems.
\item  Apply a Laplace transformation to exterior problems on $\Omega^{\rm ext}$. 
\item Solve (explicitly, numerically) the ordinary differential equations in the spatial unknown $x$.
\item Allow only “outgoing” waves by selecting the decaying solution as $x\to\pm\infty$.
\item Match Dirichlet and Neumann values at the artificial boundary.
\item Apply (explicitly, numerically) the inverse Laplace transformation.
\end{enumerate}

Especially, in case of the Dirac equation this recipe above was applied
by Hammer,  W. P\"otz and Arnold \cite{Hammer2014} in order to derive (continuous) TBCs, cf.~Eq.~\eqref{eq13}. 

The early treatment of the problem of transparent boundary conditions in
terms of the acoustic wave equation dates back to the pioneering work of
Engquist and Majda~\ci{EM1,EM2}. The further development of the concept of TBCs
for some parabolic and hyperbolic PDEs was presented by Halpern~\ci{Halpern} and
Sofronov~\ci{Sofronov}. A more comprehensive study of TBCs for the Schr\"odinger
equation and other wave equation can be found in
\cite{Arnold1998,Ehrhardt1999,Ehrhardt2001,Ehrhardt2002,Arnold2003,Jiang2004,Antoine2008,Ehrhardt2008,Sumichrast2009,Antoine2009,Ehrhardt2010,Klein2011,Arnold2012,Feshchenko2013,Antoine2014,Petrov2016}.
Earlier, approximate TBCs for the linear Schr\"odinger equation were formulated
by Shibata \cite{Shibata} and Kuska \cite{Kuska}, where the dispersion relations
for the plane waves are approximated in order to derive approximative TBCs. A
strict mathematical analysis of TBCs for different wave equations, including the
quantum mechanical Schr\"{o}dinger equation can be found in the
Refs.~\cite{Ehrhardt2002,Arnold2003,Ehrhardt2008,Antoine2014,Petrov2016}. Here
we briefly recall the formulation of TBCs on a line following the
Refs.~\cite{Ehrhardt1999,Ehrhardt2001,Ehrhardt2002,Arnold2003}.

Before proceeding to transparent quantum graphs, following the
Ref.~\ci{Hammer2014}, we briefly present the concept of TBCs
for the Dirac equation. We consider the following Dirac equation (in
units $\hbar=c=1$) posed on an interval $[0,L]\subset\mathbb{R}$:
\begin{equation}\label{eq1}
    \begin{split}
        i\partial_t \phi&=-i\partial_x \chi + m\phi,\\
        i\partial_t \chi&=-i\partial_x \phi - m\chi.
    \end{split}
\end{equation}
The initial conditions for Eq.~\eqref{eq1} are imposed as
\begin{equation}\label{eq2}
    \phi(x,0)= \phi^I(x),\quad \chi(x,0)= \chi^I(x), \quad x \in [0,L].
\end{equation}
Further, we divide the entire space into interior ($0<x<L$) and two exterior
($x\le0$ and $x\ge L$) domains.
Then, the interior problem ($x\in(0,L)$, $t>0$) reads
\begin{equation}\label{eq3}
    \begin{split}
  i\partial_t \phi &= -i\partial_x \chi + m \phi,\\
  i\partial_t \chi &= -i\partial_x \phi - m \chi,\\
  \phi(x,0) &= \phi^{I}(x),\\
  \chi(x,0) &= \chi^{I}(x),\\
  (T_0\phi)(0,t) &= \chi(0,t),\\
  (T_L\phi)(L,t) &= \chi(L,t),
 \end{split}
\end{equation}
where $T_0$ and $T_L$ denote the Dirichlet–to–Neumann (DtN) maps at the boundaries, and they are found by solving the exterior problems ($x\le0$ and $x\ge L$, $t>0$) given as
\begin{equation}\label{eq4}
    \begin{split}
i\partial_t \phi &= -i\partial_x \chi + m \phi,\\
i\partial_t \chi &= -i\partial_x \phi - m \chi,\\
\phi(x,0) &= 0,\\
\chi(x,0) &= 0,\\
\phi(0,t) &= \Phi_1(t),\\
\chi(0,t) &= (T_0\phi)(0,t),\\
\phi(L,t) &= \Phi_2(t),\\
\chi(L,t) &= (T_L\phi)(L,t).
 \end{split}
\end{equation}
By introducing the following Laplace transformations
\begin{align}
    \widetilde{\phi}(x,s)&= \int_{0}^{+\infty} \phi(x,t)\,e^{-st}\, dt,\label{eq5}\\
   \widetilde{\chi}(x,s)&= \int_{0}^{+\infty} \chi(x,t)\,e^{-st}\, dt,\label{eq6}
\end{align}
one can rewrite Eq.~\eqref{eq1} as ordinary differential equations
\begin{equation}\label{eq7}
    \begin{split}
-i\partial_{x} \widetilde{\chi}(x,s) &= (is-m) \widetilde{\phi}(x,s),\\
-i\partial_{x} \widetilde{\phi}(x,s) &= (is+m) \widetilde{\chi}(x,s).
 \end{split}
\end{equation}
The general solution of the system \eqref{eq7} can be written as
\begin{equation}\label{eq8}
    \begin{split}
&\widetilde{\phi}(x,s) = c_1 \,e^{-\sqrt[+]{s^2+m^2}x} + c_2 \,e^{\sqrt[+]{s^2+m^2}x},\\
&\widetilde{\chi}(x,s) = -c_1\kappa \,e^{-\sqrt[+]{s^2+m^2}x} + c_2\kappa
\,e^{\sqrt[+]{s^2+m^2}x},
 \end{split}
\end{equation}
where $\kappa=\frac{\sqrt[+]{is-m}}{\sqrt[+]{is+m}}$.
Since we require $\widetilde{\phi}$, $\widetilde{\chi}\in L_2(0,+\infty)$,
we obtain for $x\ge L$ (right exterior domain) the condition $c_1=0$,
and thus
\begin{equation}\label{eq9}
    \begin{split}
      \widetilde{\phi}(x,s) &= c_2 \,e^{\sqrt[+]{s^2+m^2} x},\\
      \widetilde{\chi}(x,s) &= c_2\kappa \,e^{\sqrt[+]{s^2+m^2}x}.
 \end{split}
\end{equation}
From the initial condition of the problem \eqref{eq4} we obtain
\begin{equation*}
   c_2=\widetilde{\phi}(L,s)e^{-\sqrt[+]{s^2+m^2} L} = \widetilde{\Phi}_2(s)e^{-\sqrt[+]{s^2+m^2} L},
\end{equation*}
hence
\begin{equation}\label{eq10}
    \begin{split}
       \widetilde{\phi}(x,s) &= \widetilde{\Phi}_2(s) \,e^{\sqrt[+]{s^2+m^2} (x-L)},\\
       \widetilde{\chi}(x,s) &=\widetilde{\Phi}_2(s)\kappa \,e^{\sqrt[+]{s^2+m^2} (x-L)} =
\kappa\widetilde{\phi}(x,s).
\end{split}
\end{equation}
Thus, one gets finally
\begin{equation}\label{eq11}
\widetilde{\chi}(L,s) =\kappa\widetilde{\phi}(L,s).
\end{equation}
Using the inverse Laplace transformation we have
\begin{equation}\label{eq12}
    \begin{split}
       \chi(L,t)=&\frac{d}{dt}\int_0^t I_0\bigl(m(t-\tau)\bigr)\phi(L,\tau)\,d\tau\\
        &+im\int_0^t I_0\bigl(m(t-\tau)\bigr)\phi(L,\tau)\,d\tau,
\end{split}
\end{equation}
where $I_0(z)$ denotes the modified Bessel function. Analogously, one can get the
TBC at $x=0$ \ci{Hammer2014}:
\begin{equation}\label{eq13}
    \begin{split}
         \chi(0,t)=&-\frac{d}{dt}\int_{0}^{t}I_0\bigl(m(t-\tau)\bigr)\phi(0,\tau)\,d\tau\\
          &-im\int_{0}^{t}I_0\bigl(m(t-\tau)\bigr)\phi(0,\tau)\,d\tau.
    \end{split}
\end{equation}

In this paper we extend the above procedure to the case of Dirac particle
motion on quantum graphs. We note that so far the practical applications of the
TBC concept for physical systems is quite restricted, except, may be the
Refs.~\ci{tbcqg,tbckuska,tbcnlse}, where linear and nonlinear (soliton) wave
dynamics in transparent graphs have been considered.

\begin{figure}[t!]
  \includegraphics[width=70mm]{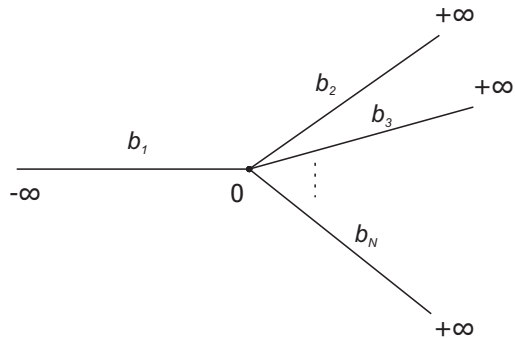}
   \caption{Star graph with $N$ semi-infinite bonds.} \label{pic1}
\end{figure}

\section{Dirac equation on quantum graphs}

The Dirac equation on quantum graphs has been considered first in \ci{Bulla1990}. 
Later, a strict treatment of the problem with the derivation of general self-adjoint vertex boundary conditions and a scattering approach was presented in \ci{Bolte}.

Here, following the Ref.~\ci{Bolte}, we briefly recall the modelling
of relativistic Dirac particles in quantum graphs. 
For nonrelativistic
particles, the quantum graph concept has been introduced first by Exner, Seba
and Stovicek~\cite{Exner1} to describe free quantum motion on branched wires.
In fact, the early treatment of the quantum mechanical motion in branched
molecular structures dates back to the Refs.~\cite{Pauling,Rud,Alex}, who
studied it in the context of organic molecules. 
Later, Kostrykin and Schrader~\cite{Kost} derived the general boundary conditions providing
self-adjointness of the Schr\"odinger operator on graphs. Relativistic quantum
mechanics described by Dirac~\cite{Bolte} and Bogoliubov-de Gennes
operators~\cite{KarimBdG} on graphs have been studied recently. Parity-time
(PT) symmetric quantum graphs are considered in \ci{Jambul}. 
Hul \emph{et al.}~\cite{Hul} considered experimental realization of quantum graphs in
optical microwave networks. An important topic related to quantum graphs was
studied in the context of quantum chaos theory and spectral statistics
\cite{Uzy1,Uzy2,Bolte}. Spectral properties and the band structure of periodic
quantum graphs also attracted much interest \cite{Exner15,Grisha}. 
The possibility of tuning the directed transport in quantum driven star graph has
been also studied in \cite{yusupov2015}, where the reflection at the central
vertex has been discussed.

Thus, quantum graphs can be determined as one- or quasi-one dimensional
branched quantum wires, where the particle and wave motion can be described in
terms of quantum mechanical wave equations on metric graphs by imposing the
boundary conditions at the branching points (vertices) and bond ends. 
The metric graph itself is a set of bonds with assigned length and which are
connected to each other at the vertices according to a rule called the topology
of a graph, which is given in terms of the adjacency matrix~\cite{Uzy1,Uzy2}:
\begin{equation*}
C_{ij}=C_{ji}=
      \begin{cases}
          1 & \text{ if }\; i\; \text{ and }\; j\; \text{ are connected, }\\
          0 & \text{ otherwise, }
      \end{cases}
\end{equation*}
for $i,j=1,2,\dots,N$.

We start with, although the simplest, but very important, star-shaped graph
shown in Fig.~\ref{pic1}. Star graphs can be considered as the smallest
building blocks for any more complicated graphs. The Dirac equation (in units
$\hbar=c=1$) on each bond ($b_j$) of a star graph can be written as
\begin{equation}\label{eq14}
    \begin{split}
        i\partial_t\phi_j&=-i\partial_x\chi_j+m\phi_j,\\
         i\partial_t\chi_j&=-i\partial_x\phi_j-m\chi_j.
    \end{split}
\end{equation}

In terms of the Dirac operator $D$ the system of equations~\eqref{eq14} can be
written as
\begin{equation}\label{eq15}
     i\partial_t\psi_j=D\psi_j,
\end{equation}
where $\psi_j=(\phi_j, \chi_j)^\top$, and
\begin{equation}\label{eq16}
    D:=-i\sigma_x\partial_x+m\sigma_z,
\end{equation}
with the Pauli matrices
$\sigma_x=\begin{pmatrix}0&1\\1&0\end{pmatrix}$,
$\sigma_z=\begin{pmatrix}1&0\\0&-1\end{pmatrix}$.

To solve Eq.~\eqref{eq14}, one needs to impose self-adjoint boundary conditions
at the vertex (branching point). To do this, we introduce the following scalar
product on the star graph as
\begin{equation}\label{eq17}
   \langle\psi,\varphi\rangle
   =\sum_{j=1}^N\int_{b_j} \bigl(\phi_j u_j^*+\chi_j v_j^*\bigr)\,dx,
\end{equation}
where
$\psi=(\psi_1,\psi_2,\dots,\psi_N)$, $\varphi=(\varphi_1,\varphi_2,\dots,\varphi_N)$,
and $\psi_j=(\phi_j,\chi_j)^\top$, $\varphi_j=(u_j, v_j)^\top$
for $j=1,2,\dots,N$.

Defining the following skew-Hermitian scalar product on a graph
\begin{equation}\label{eq18}
    \begin{split}
   \Omega(\psi,\varphi)&=\langle D\psi,\varphi\rangle-\langle\psi,D\varphi\rangle\\
    &=i\sum_{j=1}^N \bigl[\phi_j(0)v_j(0)-\phi_j(L_j)v_j(L_j)\\
    &\quad+\chi_j(0)u_j(0)-\chi_j(L_j)u_j(L_j)\bigr].
\end{split}
\end{equation}
One can prove (see, \ci{Bolte}) that self-adjointness of the Dirac operator on
the graph is provided by the condition $\Omega(\psi,\varphi)=0$. A set of
vertex boundary conditions fulfilling this condition can be written as
\begin{equation}\label{eq19}
    \begin{split}
    \phi_1(0)&=\phi_2(0)=\dots=\phi_N(0),\\
    \chi_1(0)&+\chi_2(0)+\dots+\chi_N(0)=0,\\
     \phi_j(L_j)&=0,\quad j=1,2,\dots,N.
\end{split}
\end{equation}
The first condition represents the continuity at the vertex, while the second
one provides the Kirchhoff rule at the vertex.

\section{Dirac particles in transparent quantum graphs}

In this section we apply the procedure presented in Section~II for the derivation of
transparent vertex boundary conditions for the Dirac equation on quantum graphs. 
It should be noted that the reflectionless motion of waves and
particles in networks has been considered earlier within the different
approaches in the Refs.~\cite{Uzy07,Joly,Naim,Exner10,Kurasov,Cheon}. In
\cite{Uzy07} the S-matrix based approach is utilized for the construction of
scattering matrices providing the absence of backscattering. 
Indirect observations of the possibility for the wave transmission through the graph
vertex were presented earlier in several studies, in particular by Naimark and Solomyak~\cite{Naim}, 
using some properties of the weighted Laplacian on metric graphs. 
The possibility for reducing a homogeneous tree graph investigation,
using its symmetry, into a family of one-dimensional problems with point
interactions was considered by Exner and Lipovsky~\cite{Exner10}. 

The existence of the reflectionless and equitransmitting vertex couplings has been studied in \cite{Kurasov,Cheon}. 
Despite a certain number of publications containing a study
of reflectionless wave transmission through the graph vertices, all these
studies did not use the concept of transparent boundary conditions for the
Schr\"odinger equation. 
In \cite{Joly}, the transparent boundary conditions (TBCs) are
considered for the wave equation on a metric tree graph with self-similar
structure at infinity, where the properties of reflectionless transport are
studied numerically. 
However, the latter mentioned leads to very complicated
numerical work, whose accuracy cannot be easily controlled. 
It should be noted that the direct application of TBCs for quantum graphs converts the problem
into a very complicated numerical problem with almost uncontrolled accuracy.

Therefore, one needs to develop methods allowing to avoid the direct
utilization of a TBC at the branching point, but instead turn the vertex
boundary conditions equivalent to the TBCs. Here we combine the above concepts of
TBCs and quantum graphs to design the vertex boundary conditions providing
reflectionless transmission of waves through the graph branching points. In the
following, such graphs will be called ``transparent quantum graphs''. Recently,
such an approach has been proposed for quantum graphs described by a (linear)
Schr\"odinger equation on metric graphs~\ci{tbcqg}. 
Constraints applying the usual continuity and the Kirchhoff rule to the 
TBC at the vertex were found in the form of simple sum rule. 
An extension to the case of solitons described by the
nonlinear Schr\"odinger equation on metric graphs has been presented in
\ci{tbcnlse}. 
Here we will use a similar approach for relativistic
quasiparticles in networks modeled by the Dirac equation on quantum graphs.
Such quasiparticles appear, e.g., in branched graphene nanoribbons (see,
Refs.~\ci{Andriotis,Chen,Xu,Kvashnin}).

Let us consider a star graph shown in Fig.~\ref{pic1}. For simplicity (without
losing generality) we consider first a star graph with three bonds. 
To each bond $b_j$ of the graph we assign a coordinate $x_j$, which indicates the
position along the bond: for bond $b_1$ it is $x_1\in (-\infty,0]$ and for
$b_{1,2}$ they are $x_{2,3}\in [0,+\infty)$. In the following we use the
shorthand notation $\Psi_j(x)$ for $\Psi_j(x_j)$ and it is understood that $x$
is the coordinate on the bond $j$ to which the component $\Psi_j$ refers. The
Dirac equation (in units $\hbar=c=1$) on each bond $b_{j}$ $(b_{1}~(-\infty;
0], b_{2,3}~[0; +\infty) )$ of such graph can be written as
\begin{equation}\label{eq20}
    \begin{split}
   i\partial_{t} \phi_{j}&=-i \partial_{x} \chi_{j} + m \phi_{j},\\
   i\partial_{t} \chi_{j}&=-i \partial_{x} \phi_{j} - m \chi_{j}.
\end{split}
\end{equation}
We impose the boundary conditions at the vertex as follows
\begin{align}
  \alpha_{1} \phi_{1}|_{x=0} &= \alpha_{2} \phi_{2}|_{x=0} =
   \alpha_{3} \phi_{3}|_{x=0},\label{eq21}\\
   \frac{1}{\alpha_{1}} \chi_{1}|_{x=0} &= \frac{1}{\alpha_{2}} \chi_{2}|_{x=0} +
   \frac{1}{\alpha_{3}} \chi_{3}|_{x=0}.\label{eq22}
\end{align}

We note that the vertex boundary conditions given by 
Eqs.~\eqref{eq21} and \eqref{eq22} are self-adjoint, as they are consistent with the condition $\Omega(\psi,\varphi)=0$, where $\Omega(\psi,\varphi)$ is determined by Eq.~\eqref{eq18}.

Our task is to derive (self-adjoint) transparent vertex boundary conditions (VBCs) using the same procedure as in Section~II and finding the constraints, which make the VBCs in Eqs.~\eqref{eq21} and ~\eqref{eq22} equivalent to the transparent ones. Such boundary conditions provide reflectionless transmission of
Dirac quasiparticles at the vertex. In analogy with the procedure described in Section~II, the
`interior' problem for the star graph is given on the bond $b_1$ and
can be written as

\begin{equation}\label{eq23}
    \begin{split}
       i\partial_{t} \phi_{1} &= -i \partial_{x} \chi_{1} + m \phi_{1},\\
       i\partial_{t} \chi_{1} &= -i \partial_{x} \phi_{1} - m \chi_{1},\\
         \phi_{1}(x,0) &= \phi_{1}^{I}(x),\\
           \chi_{1}(x,0) &= \chi_{1}^{I}(x),\\
         (T\phi_{1})(0,t) &= \chi_{1}(0,t).
\end{split}
\end{equation}
The two `exterior' problems on $b_{2,3}$ are given as
\begin{equation}\label{eq24}
    \begin{split}
    i\partial_{t} \phi_{2,3} &= -i \partial_{x} \chi_{2,3} + m \phi_{2,3},\\
    i\partial_{t} \chi_{2,3} &= -i \partial_{x} \phi_{2,3} - m \chi_{2,3},\\
    \phi_{2,3}(x,0) &= 0,\\
    \chi_{2,3}(x,0) &= 0,\\
    \phi_{2,3}(0,t) &= \Phi_{2,3}(t),\\
     \chi_{2,3}(0,t) &= (T \phi_{2,3})(0,t).
\end{split}
\end{equation}
Further, we introduce the following Laplace transformation:
\begin{align}
   \widetilde{\phi}_{2,3}(x,s)&= \int_{0}^{+\infty} \phi_{2,3}(x,t)\,e^{-st}\,dt,\label{eq25}\\
   \widetilde{\chi}_{2,3}(x,s)&= \int_{0}^{+\infty} \chi_{2,3}(x,t)\,e^{-st}\,dt.\label{eq26}
\end{align}

Then, the two `exterior' problems \eqref{eq24} are written as
\begin{equation}\label{eq27}
    \begin{split}
       -i \partial_{x} \widetilde{\chi}_{2,3}(x,s) &= (is-m) \widetilde{\phi}_{2,3}(x,s),\\
       -i \partial_{x} \widetilde{\phi}_{2,3}(x,s) &= (is+m)\widetilde{\chi}_{2,3}(x,s).
\end{split}
\end{equation}
The general solution of the system \eqref{eq27} reads
\begin{equation}\label{eq28}
    \begin{split}
 \widetilde{\phi}_{2,3}(x,s) &= c_{2,3}^{(1)}\, e^{-\sqrt[+]{s^2+m^2} x}
 + c_{2,3}^{(2)} \,e^{\sqrt[+]{s^2+m^2} x},\\
 \widetilde{\chi}_{2,3}(x,s) &= - c_{2,3}^{(1)} \kappa \,e^{-\sqrt[+]{s^2+m^2} x}
+ c_{2,3}^{(2)}\kappa\, e^{\sqrt[+]{s^2+m^2} x},
\end{split}
\end{equation}
where $\kappa=\frac{\sqrt[+]{is-m}}{\sqrt[+]{is+m}}$.
Since $\widetilde{\phi}_{2,3}$, $\widetilde{\chi}_{2,3}\in L_2(0;+\infty)$, we obtain
for the `exterior' problems on $b_{2,3}$
\begin{equation*}
     c_{2,3}^{(1)}=0,
\end{equation*}
hence the exterior solution is
\begin{equation}\label{eq29}
    \begin{split}
      \widetilde{\phi}_{2,3}(x,s) &= c_{2,3}^{(2)} \,e^{\sqrt[+]{s^2+m^2} x},\\
      \widetilde{\chi}_{2,3}(x,s) &= c_{2,3}^{(2)}\kappa \,e^{\sqrt[+]{s^2+m^2}
x}.
   \end{split}
\end{equation}
Now, the initial conditions of the problem \eqref{eq24} yield
\begin{equation*}
   c_{2,3}^{(2)} =\widetilde{\phi}_{2,3}(0,s) =
   \widetilde{\Phi}_{2,3}(s),
\end{equation*}
hence
\begin{equation}\label{eq30}
    \begin{split}
   \widetilde{\phi}_{2,3}(x,s) &= \widetilde{\Phi}_{2,3}(s)\, e^{\sqrt[+]{s^2+m^2} x},\\
     \widetilde{\chi}_{2,3}(x,s) &= \widetilde{\Phi}_{2,3}(s)\kappa\,
    e^{\sqrt[+]{s^2+m^2} x}\\
      &= \kappa \widetilde{\phi}_{2,3}(x,s).
    \end{split}
\end{equation}

From the vertex boundary conditions \eqref{eq21}-\eqref{eq22} we get
\begin{equation}\label{eq31}
    \begin{split}
\widetilde{\phi}_{2,3}|_{x=0} &= \frac{\alpha_{1}}{\alpha_{2,3}}
\widetilde{\phi}_{1}|_{x=0},\\
\widetilde{\chi}_{2,3}|_{x=0} &=\kappa\frac{\alpha_{1}}{\alpha_{2,3}} \widetilde{\phi}_{1}|_{x=0},\\
\widetilde{\chi}_{1}|_{x=0} &= \frac{\alpha_{1}}{\alpha_{2}}
\widetilde{\chi}_{2}|_{x=0} + \frac{\alpha_{1}}{\alpha_{3}}
\widetilde{\chi}_{3}|_{x=0} \\
&= \frac{\alpha_{1}}{\alpha_{2}} \Bigl( \kappa \frac{\alpha_{1}}{\alpha_{2}}
\widetilde{\phi}_{1}|_{x=0} \Bigr)
+ \frac{\alpha_{1}}{\alpha_{3}} \Bigl( \kappa\frac{\alpha_{1}}{\alpha_{3}}\widetilde{\phi}_{1}|_{x=0}\Bigr) \\
&= \alpha_{1}^{2} \Bigl( \frac{1}{\alpha_{2}^{2}} + \frac{1}{\alpha_{3}^{2}}
\Bigr) \kappa \widetilde{\phi}_{1}|_{x=0}.
 \end{split}
\end{equation}
The Laplace transformed Kirchhoff rule in Eq.~\eqref{eq22} yields
\begin{multline}\label{eq32}
     \chi_1(0,t)=A\biggl[\frac{d}{dt}\int_0^t I_0\bigl(m(t-\tau)\bigr)\phi_1(0,\tau)\,d\tau\\
               \quad +im\int_0^t I_0\bigl(m(t-\tau)\bigr)\phi_1(0,\tau)\,d\tau\biggr],
\end{multline}
where $A=\alpha_1^2\bigl(\frac{1}{\alpha_2^2}+\frac{1}{\alpha_3^2}\bigr)$ and
$I_0(z)$ denotes the modified Bessel function.

\begin{figure}[t!]
\includegraphics[width=65mm]{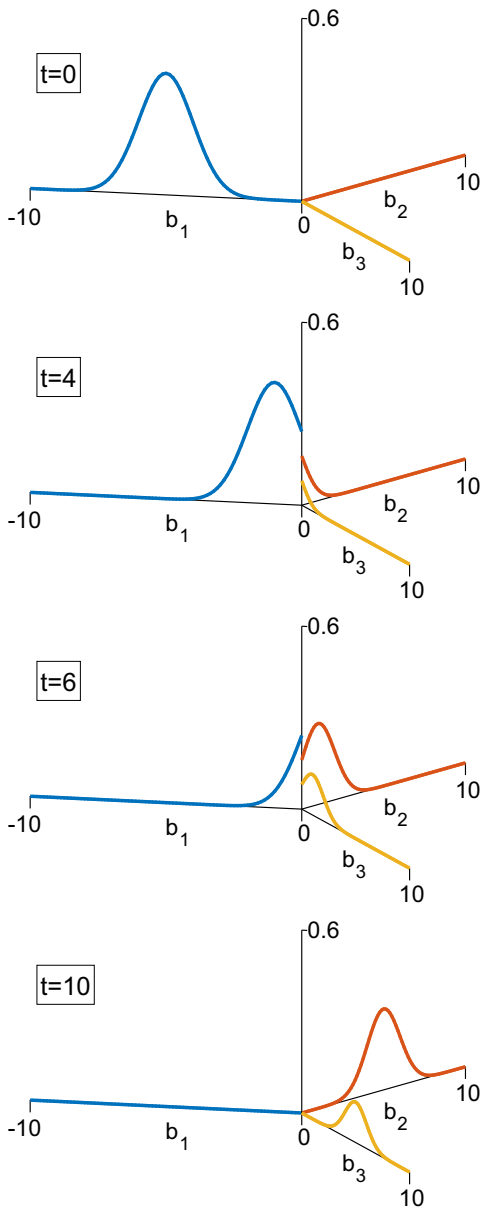}
\caption{(color online). The position probability density
$|\phi_j(x,t)|^2+|\chi_j(x,t)|^2$ plotted at different time moments for the
regime when the sum rule is fulfilled (no reflection occurred):
$\alpha_1=\sqrt{2/3}$, $\alpha_2=1$ and $\alpha_3=\sqrt{2}$.} \label{pic2}
\end{figure}

It is clear that the boundary condition in Eq.~\eqref{eq32} coincides with that
in Eq.~\eqref{eq12} and hence, provides reflectionless transmission of Dirac
quasiparticles for the bond $b_1$, when $A=1$, i.e. the following sum rule is
fulfilled:
\begin{equation}\label{sumrule}
     \frac{1}{\alpha_{1}^{2}} =
          \frac{1}{\alpha_{2}^{2}} + \frac{1}{\alpha_{3}^{2}}.
\end{equation}

The transparent boundary conditions given by Eq.~\eqref{eq32} do not destroy the self-adjointness of the problem,
since they are derived from the vertex boundary conditions \eqref{eq21} and \eqref{eq22} providing self-adjointness
of the Dirac operator on the graph. We emphasize the fact 
that although the boundary conditions \eqref{eq32} are derived from those in Eqs.~\eqref{eq21} and \eqref{eq22}, a derivation in the opposite direction is not possible. 

In this way, the vertex boundary conditions given by
Eqs.~\eqref{eq21}-\eqref{eq22} become equivalent to the transparent vertex
boundary conditions, provided the sum rule in Eq.~\eqref{sumrule} is fulfilled.

We note that all the above results hold true for the special massless case, $m=0$, too. 
This is the case appearing, e.g., with the Dirac quasiparticles in graphene.

In addition, the obtained sum rule can easily be generalized for any star graph with $N$ bonds.
For this general case, 
when the bond $b_1$ is considered as incoming bond, the sum rule takes form 
\begin{equation}\label{gen_sumrule}
    \alpha_1^{-2} = \sum_{j=2}^N \alpha_j^{-2}
\end{equation}
and the transmission of the wave (quasiparticle) in $N-1$ bonds becomes reflectionless. 
Furthermore, the approach can be extended for an arbitrary graph topology, having an arbitrary number of vertices, 
i.e.\ the TBC can be derived for an  arbitrary metric graph. 
This can be easily done in an analogous way as it was shown in the Ref.~\ci{tbcqg},
where for example the approach was extended for a tree graph.

\begin{figure}[t!]
\includegraphics[width=85mm]{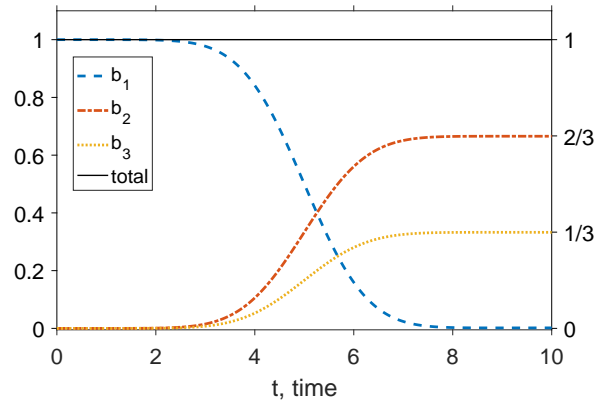}
\caption{(color online). Time dependence of the partial and total norms for the
case shown in Fig.~\ref{pic2}.} \label{pic3}
\end{figure}

\section{Numerical Experiment}

This section provides the numerical justification of our obtained results. The
configuration of the experimental set-up consists of a star graph with three
bonds. The free time evolution of the Gaussian Dirac spinor
\begin{equation*}
    G(x)=\Bigl(\frac{1}{2\pi\sigma^2}\Bigr)^{1/4}
             \exp\Bigl( -\frac{(x-x_0)^2}{4\sigma^2} \Bigr)
  \begin{pmatrix}
    1 \\1
  \end{pmatrix}
\end{equation*}
compactly supported in the first bond is studied. This is done by considering
the following initial conditions:
\begin{equation*}
     \begin{pmatrix}\phi_1(x,0) \\ \chi_1(x,0) \end{pmatrix}=G(x),
\end{equation*}
with $x_0=-5$, $\sigma=0.9$, and
\begin{equation*}
   \begin{pmatrix} \phi_{2,3}(x,0) \\ \chi_{2,3}(x,0) \end{pmatrix}
    =\begin{pmatrix} 0 \\ 0 \end{pmatrix}.
\end{equation*}
We choose $m=0.01$ such that the transmission process is rather demonstrable,
that is, the wave packet retains its shape for the considered time period. 
The leap-frog scheme (see, \ci{Hammer2014} for details) with the space
discretization $\Delta x = 0.0125$ and the time step $\Delta t = 0.01$ is
utilized for the numerical experiment. 
The plots in Fig.~\ref{pic2} show the
position probability density $|\phi_j(x,t)|^2+|\chi_j(x,t)|^2$, $j=1,2,3$ at
four consecutive time steps. 
It is clear from Fig.~\ref{pic2} that the wave
entirely transmits to the second and third bonds without any reflections when time elapses. 
We note that the above boundary conditions provide the conservation
of the total norm, which is defined as the sum of partial norms for each bond.
The time-dependence of the partial and total norms for this case is shown in
Fig.~\ref{pic3}.

\begin{figure}[t!]
\includegraphics[width=90mm]{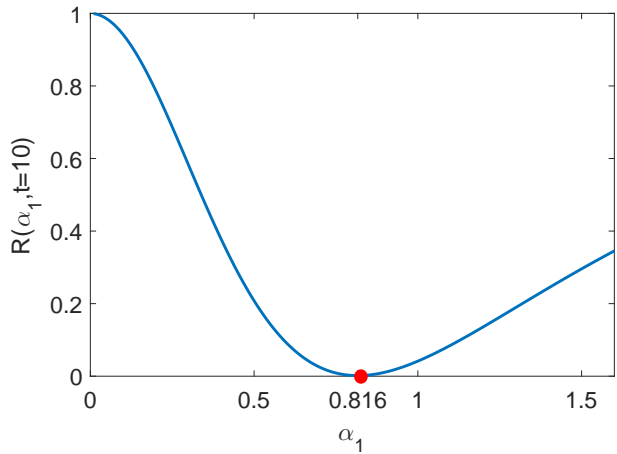}
\caption{(color online). Dependence of the vertex reflection coefficient $R$ on
the parameter $\alpha_1$ when the wave packet splitting time elapses ($t=10$).
For fixed $\alpha_2=1$ and $\alpha_3=\sqrt{2}$, $R=0$ when
$\alpha_1=\sqrt{2/3}\approx0.816$ (red dot).} \label{pic4}
\end{figure}

In Fig.~\ref{pic4} the reflection coefficient $R$ determined as the ratio of the
partial norm for the first bond to the total norm
\begin{equation*}
     R=\frac{N_1}{N_1+N_2+N_3}
\end{equation*}
is plotted as a function of $\alpha_1$ for the fixed values of $\alpha_2$ and
$\alpha_3$. From this plot one can conclude that the reflection coefficient
becomes zero only at the value of $\alpha_1$ that provides the fulfillment of
the sum rule~\eqref{sumrule}. 
This observation clearly confirms once more that the sum rule in
Eq.~\eqref{sumrule} turns the vertex boundary conditions in
Eqs.~\eqref{eq21}-\eqref{eq22} equivalent to the transparent ones.

It should also be noted here that the possibility of tuning the transition of a
wave packet through the vertex comes from selecting the proper parameters
$\alpha_2$ and $\alpha_3$ such that the `masses' of fractions will be
$N_2=1-\alpha_2^2/(\alpha_2^2+\alpha_3^2)$ and
$N_3=1-\alpha_3^2/(\alpha_2^2+\alpha_3^2)$, accordingly.

Let us note that Hammer, P\"otz and Arnold \cite{Hammer2014} also introduced so-called discrete TBCs that were analogously derived, but for the discretized Dirac equation in 1D (using a leap-frog scheme).
Finally, an ``energy'' functional 
\begin{multline}\label{eq:En}
  E^n=  \rVert \phi^{n+1/2} \rVert^2 + \rVert\chi^{n+1}\rVert^2 \\
  +\frac{\Delta t}{\Delta x} 
  \Re \bigl[(D\phi^{n+1/2},\chi^{n+1})\bigr]=\text{const}=E^0,
\end{multline}
with the symmetric spatial difference operator
\begin{equation*}
  (D\phi^{n+1/2})_{j-1/2}=\phi^{n+1/2}_j-\phi^{n+1/2}_{j-1},
\end{equation*}
was identified, which is exactly conserved by the leap-frog scheme, 
even in the presence of non-constant mass and
potential terms.

\section{Conclusions}

In this paper we studied the problem of a Dirac particle dynamics in
transparent quantum graphs. These latter mentioned are determined as branched
quantum wires providing reflectionless transmission of waves at the branching
points. The boundary conditions for the time-dependent Dirac equation on
graphs, providing absence of backscattering at the vertex are formulated
explicitly. A constraint that makes the usual Kirchhoff-type boundary
conditions at the vertex equivalent to those of transparent ones is derived in
the form of a simple sum rule.

A reflectionless transmission of the Gaussian wave packet through the vertices,
provided these constraints are fulfilled, is shown numerically for the star
graph. This approach can be directly extended for arbitrary graph topologies,
which contain any subgraph connected to two or more outgoing, semi-infinite
bonds.

The proposed model can be applied for a broad range of practically important
problems in condensed matter physics, where Dirac quasiparticles appear in
branched graphene nanoribbons, CNT networks, and branched topological
insulators. Moreover, the approach can be utilized for modeling polarons
dynamics~\ci{Campbell1981,Campbell1982} in branched conducting polymers, where
reflectionless transmission of charge carriers through the branching points can
occur.


\end{document}